\documentclass{prop2015}% no class options needed by now
\usepackage[english]{babel}
\keywords{Population dynamics, nonequilibrium many-body problem, thermalization, exact results, long-time limit.}
\title{The role of average time dependence on the relaxation of excited electron populations in nonequilibrium many-body physics}
\author[A. F. Kemper]{A. F. Kemper\inst{1,}}
\author[H. R. Krishnamurthy]{H. R. Krishnamurthy\inst{2}}
\author[J. K. Freericks]{J. K. Freericks\inst{3},\footnote{Corresponding author\quad E-mail:~\textsf{james.freericks@georgetown.edu}}}
\address[1]{Department of Physics, North Carolina State University, Raleigh, North Carolina 27695, USA}
\address[2]{Department of Physics, Indian Institute of Science, Bangalore-560012, India}
\address[3]{Department of Physics, Georgetown University, 37th and O Sts., NW, Washington, District of Columbia, 20057-1789 USA}
\shortauthors{A. F Kemper {\it et al.}}
\begin{abstract}
We examine the exact equation of motion for the relaxation of populations of strongly correlated electrons  after a nonequilibrium excitation by a pulsed field, and prove that the populations do not change when the Green's functions have no average time dependence. We show how the average time dependence enters into the equation of motion to lowest order and describe what governs the relaxation process of the electron populations in the long-time limit. While this result may appear, on the surface, to be required by any steady-state solution, the proof is nontrivial, and provides new critical insight into how nonequilibrium populations relax, which goes beyond the assumption that they thermalize via a simple relaxation rate determined by the imaginary part of the self-energy, or that they can be described by a quasi-equilibrium condition with a Fermi-Dirac distribution and a time-dependent temperature. We also discuss the implications of this result to approximate theories, which may not satisfy the exact relation in the equation of motion.
\end{abstract}
\shortabstract
\begin{document}
\maketitle
% \noindent

\section{Introduction}

Recently, there have been many experiments in strongly correlated electron materials that follow the pump-probe paradigm---namely, the system is excited with an ultrafast and ultra-intense electromagnetic field and then the system is probed (often with weaker pulses) after different time delays from the pump pulse. One area of interest to examine is the so-called population dynamics, which examines how the expectation value of the density of the electrons ($\langle n_{\bf k}(t)\rangle$, also called the population) changes as functions of momentum and time~\cite{lanzara,shen,rameau}; we are especially interested in how the system approaches a thermal state in the long time limit. 

In order to describe such systems theoretically, we need to work with two-time Green's functions defined on the Kadanoff-Baym-Keldysh contour~\cite{kadanoff_baym,keldysh}, which starts at some initial time ($t_{min}$), runs along the real axis to some maximum time ($t_{max}$), then returns along the real axis to the initial time, and finally runs a distance equal to the inverse temperature (of the initial equilibrium state, $\beta$) parallel to the negative imaginary axis. It is customary to define a contour-ordered Green's function with each time argument lying on the contour and then determine the equation of motion by differentiating with respect to each time $t$ and $t'$. Details for how to do this appear in many different sources, including recent texts~\cite{noneq_text}. We do not repeat them here to save space. But we will write out the equations of motion below for the Green's functions we work with in this paper---namely, the retarded and the lesser Green's function. 

\section{Dynamics of Retarded Green's Functions}

To begin, we note that we will be working with the Wigner time coordinates~\cite{wigner} of average time $t_{ave}=(t+t')/2$ and relative time $t_{rel}=t-t'$. The retarded Green's function is given by
\begin{equation}
G^R_{\bf k\sigma}(t,t')=-i\theta(t-t')\langle \{c_{\bf k\sigma}^{\phantom\dagger}(t),c_{\bf k\sigma}^\dagger(t')\}_+\rangle
\end{equation}
where the angle brackets $\langle\cdots\rangle$ denote the trace over all states weighted by the initial equilibrium density matrix given by $\exp[-\beta\mathcal{H}(t_{min})]/\mathcal{Z}$, where $\mathcal{Z}={\rm Tr} \exp[-\beta\mathcal{H}(t_{min})]$ is the partition function and $\mathcal{H}(t)$ is the time-dependent Hamiltonian. The $\theta(t)$ function is the unit step function, $c_{\bf k\sigma}^{\phantom\dagger}$ ($c_{\bf k\sigma}^\dagger$) are the electron annihilation (creation) operators for an electron with
momentum {\bf k} and spin $\sigma$. The symbol $O(t)$ denotes the Heisenberg representation of the operator $O$, where $O(t)=\mathcal{U}^\dagger(t,t_{min})O\mathcal{U}(t,t_{min})$, and the evolution operator is the time-ordered product $\mathcal{U}(t,t_{min})=\mathcal{T}_t \exp[-i\int_{t_{min}}^td\bar t\mathcal{H}(\bar t)]$, using standard notation. Finally, $\{\cdots,\cdots\}_+$ is the anticommutator. Since we will be working in the paramagnetic phase, we neglect the spin index below, for simplicity.

If we let $\mu$ denote the chemical potential, $\epsilon_{\bf k}$ the noninteracting bandstructure for a single-band model, introduce the electric field with a spatially uniform, but time-dependent vector potential ${\bf A}(t)$ and the Peierls substitution with ${\bf E}(t)=-\partial_t {\bf A}(t)$ (for units where $\hbar=c=e=1$), and we assume the self-energy is local, as in dynamical mean-field theory, then
the retarded Green's function, when expressed as a function of average and relative time, satisfies two 
equations of motion:
\begin{eqnarray}
\Bigr [&i&\partial_{t_{rel}}+\mu-\frac12\epsilon_{{\bf k}-{\bf A}\left (t\right )}-\frac12\epsilon_{{\bf k}-{\bf A}\left (t'\right )}\Bigr ]G_{\bf k}^R(t_{ave},t_{rel})\nonumber\\
&=&\delta(t_{rel})+\frac12\int_{t'}^t d\bar t\Sigma^R\left (\frac{t+\bar t}{2},t-\bar t\right )G_{\bf k}^R\left (\frac{\bar t+t'}{2},\bar t-t'\right )\nonumber\\
&+&\frac12\int_{t'}^t d\bar tG^R_{\bf k}\left (\frac{t+\bar t}{2},t-\bar t\right )\Sigma^R\left (\frac{\bar t+t'}{2},\bar t-t'\right )
\end{eqnarray}
and
\begin{eqnarray}
\Bigr [&i&\partial_{t_{ave}}-\frac12\epsilon_{{\bf k}-{\bf A}\left (t\right )}+\frac12\epsilon_{{\bf k}-{\bf A}\left (t'\right )}\Bigr ]G_{\bf k}^R(t_{ave},t_{rel})\nonumber\\
&=&\frac12\int_{t'}^t d\bar t\Sigma^R\left (\frac{t+\bar t}{2},t-\bar t\right )G_{\bf k}^R\left (\frac{\bar t+t'}{2},\bar t-t'\right )\nonumber\\
&-&\frac12\int_{t'}^t d\bar tG^R_{\bf k}\left (\frac{t+\bar t}{2},t-\bar t\right )\Sigma^R\left (\frac{\bar t+t'}{2},\bar t-t'\right )\,.
\end{eqnarray}
We want to use these two equations to determine the leading behavior of the retarded Green's function on $t_{ave}$ for long times, as the system approaches thermalization (or more generally, a long-time steady state). To do this, we will be taking the Fourier transformation with respect to the relative time, assume ${\bf A}(t)={\bf A}(t')=0$ since the pulse occurs at early times only, and we assume the average time dependence of the Green's function and the self-energy are both weak, so they can be expanded in a Taylor series expansion with respect to times near $t_{ave}$. Then after significant algebra, we find that the two equations for the retarded Green's function become
\begin{equation}
\left [ \omega+\mu-\epsilon_{\bf k}-\Sigma^R(t_{ave},\omega)\right ] G^R_{\bf k}(t_{ave},\omega)=1
\end{equation}
and
\begin{eqnarray}
[1&-&\partial_\omega \Sigma^R(t_{ave},\omega)]i\partial_{t_{ave}} G^R_{\bf k}(t_{ave},\omega)\nonumber\\
&=&-i\partial_{t_{ave}}\Sigma^R(t_{ave},\omega)\partial_\omega G^R_{\bf k}(t_{ave},\omega)\,.
\label{eq: eom_ret}
\end{eqnarray}
If we solve the first equation, we find
\begin{equation}
G^R_{\bf k}(t_{ave},\omega)=\frac{1}{\omega+\mu-\epsilon_{\bf k}-\Sigma^R(t_{ave},\omega)}\,,
\label{eq: gret}
\end{equation}
which also satisfies the second equation (this can be verified by direct substitution). Hence, in the long-time limit the retarded Green's function assumes its equilibrium form, with a weakly average-time-dependent self-energy. This agrees with the well-known result that the system relaxes into the retarded Green's function of the steady state before the populations of the electrons relax; i.e., the density of states rapidly assumes its long-time limit, while the distribution of the electrons within those density of states takes longer to reach the long-time limit.  The behavior of the lesser Green's function (and the populations) is much more complicated, as we will discover next.

\section{Dynamics of Excited Populations}

To begin discussing the populations, we first need to define the lesser Green's function, which satisfies
\begin{equation}
G^<_{\bf k\sigma}(t,t')=i\langle c^\dagger_{\bf k\sigma}(t')c^{\phantom\dagger}_{\bf k\sigma}(t)\rangle\,,
\end{equation}
from which, we find the population satisfies
\begin{equation}
\langle n_{\bf k}(t)\rangle=-iG^<_{\bf k\uparrow}(t,t)-iG^<_{\bf k\downarrow}(t,t)\,,
\end{equation}
where we used the fact that the density $n_{\bf k}$ is equal to $c^\dagger_{\bf k\uparrow}c^{\phantom\dagger}_{\bf k\uparrow}+c^\dagger_{\bf k\downarrow}c^{\phantom\dagger}_{\bf k\downarrow}$.

Since the population is an equal-time expectation value, it does not depend on the relative time, so we focus
on the average-time equation of motion for the lesser Green's function (suppressing the spin index again)
\begin{eqnarray}
\Bigr [&i&\partial_{t_{ave}}-\frac12\epsilon_{{\bf k}-{\bf A}\left (t\right )}+\frac12\epsilon_{{\bf k}-{\bf A}\left (t'\right )}\Bigr ]G_{\bf k}^<(t_{ave},t_{rel})\nonumber\\
&=&\int_{-\infty}^t d\bar t\Sigma^R\left (\frac{t+\bar t}{2},t-\bar t\right )G_{\bf k}^<\left (\frac{\bar t+t'}{2},\bar t-t'\right )\nonumber\\
&-&\int_{-\infty}^t d\bar tG^R_{\bf k}\left (\frac{t+\bar t}{2},t-\bar t\right )\Sigma^<\left (\frac{\bar t+t'}{2},\bar t-t'\right )\nonumber\\
&+&\int_{-\infty}^{t'} d\bar t\Sigma^<\left (\frac{t+\bar t}{2},t-\bar t\right )G_{\bf k}^A\left (\frac{\bar t+t'}{2},\bar t-t'\right )\nonumber\\
&-&\int_{-\infty}^{t'} d\bar tG^<_{\bf k}\left (\frac{t+\bar t}{2},t-\bar t\right )\Sigma^A\left (\frac{\bar t+t'}{2},\bar t-t'\right )
\,.
\label{eq: pop0}
\end{eqnarray}
In this equation, we introduced the advanced Green's function. It is
related to the retarded one via $G^A_{\bf k}(t_{ave},\omega)=G^R_{\bf k}(t_{ave},\omega)^*$, where the star denotes complex conjugation; we use this result below.

Since we care only about $t_{rel}=0$ for the populations, we set $t_{rel}=0$ ($t=t'=t_{ave}$), shift the integral over $\bar t$ to $\bar t \rightarrow \bar t +t_{ave}$, and introduce the Fourier transform with respect to the relative time in the integrands to yield
\begin{eqnarray}
&i&\partial_{t_{ave}}G_{\bf k}^<(t_{ave},t_{rel}=0)=\frac{1}{4\pi^2}\int_{-\infty}^0 d\bar t \int_{-\infty}^\infty d\omega \int_{-\infty}^\infty d\omega' \nonumber\\
&\times&e^{i(\omega-\omega')\bar t}\Biggr \{ \Sigma^R\left (t_{ave}+\frac{\bar t}{2},\omega\right )G_{\bf k}^<\left (t_{ave}+\frac{\bar t}{2},\omega'\right )\nonumber\\
&~&\quad\quad\quad\,\,\,\,\,\,-
G_{\bf k}^R\left (t_{ave}+\frac{\bar t}{2},\omega\right )\Sigma^<\left (t_{ave}+\frac{\bar t}{2},\omega'\right )\nonumber\\
&~&\quad\quad\quad\,\,\,\,\,\,+\Sigma^<\left (t_{ave}+\frac{\bar t}{2},\omega\right )G_{\bf k}^A\left (t_{ave}+\frac{\bar t}{2},\omega'\right )\nonumber\\
&~&\quad\quad\quad\,\,\,\,\,\,-
G_{\bf k}^<\left (t_{ave}+\frac{\bar t}{2},\omega\right )\Sigma^A\left (t_{ave}+\frac{\bar t}{2},\omega'\right )\Biggr \}\,.
\label{eq: pop1}
\end{eqnarray}

Note that no approximations beyond the assumption of a local self-energy went into the derivation of Eq.~(\ref{eq: pop1}), and that assumption is not required. Hence these results are exact and hold for
every strongly correlated system driven out of equilibrium.  We will see next, that if there is no average
time dependence to any of these quantities, then the populations do not change with time.

\section{Proof of the Need for Average Time Dependence to have Populations Relax}

The system starts in equilibrium, before the field is applied, and in this case, the populations are constant in time, and hence their derivative with respect to time vanishes. This result follows directly from Eq.~(\ref{eq: pop1}) as we now show. In equilibrium, the Green's functions and self-energies have no average time dependence, so all of the $\bar t$ dependence in the integral comes from the exponential term. Using
the well-known Sokhotskyi-Plemelj-Dirac identity
\begin{eqnarray}
\int_{-\infty}^0 d\bar t e^{i(\omega-\omega')\bar t} &=&\frac{1}{i} \lim_{\epsilon\rightarrow 0+}
\frac{1}{\omega-\omega'-i\epsilon}\nonumber\\
&=&\frac{1}{i}\left [\frac{\mathcal{P}}{\omega-\omega'}+i\pi\delta(\omega-\omega')\right ]\,,
\end{eqnarray}
where $\mathcal{P}$ denotes the principal value when the expression is inserted into an integral, leads to
\begin{eqnarray}
&i&\partial_{t_{ave}}G_{\bf k}^<(t_{ave},t_{rel}=0)= \nonumber\\
&~&\frac{i}{2\pi} \int_{-\infty}^\infty d\omega \Biggr \{ {\rm Im}\Sigma^R\left ( \omega\right )G_{\bf k}^<\left (\omega\right )-
{\rm Im}G_{\bf k}^R\left (\omega\right )\Sigma^<\left (\omega\right )\Biggr \}+\nonumber\\
&~&\frac{1}{2i\pi^2} \int_{-\infty}^\infty d\omega \int_{-\infty}^\infty d\omega'\Biggr \{ {\rm Re}\Sigma^R\left ( \omega\right )G_{\bf k}^<\left (\omega'\right )\nonumber\\
&~&~~~~~~~~~~~~~~~~~~~~~~~~~~~~~~~~~~~~~~~~~~~-
{\rm Re}G_{\bf k}^R\left (\omega\right )\Sigma^<\left (\omega'\right )
\Biggr \}\frac{\mathcal{P}}{\omega-\omega'}\,.
\label{eq: pop2}
\end{eqnarray}
Next, we use the Kramers-Kronig relation to simplify the double integral into a single integral. Namely, we
use
\begin{equation}
\frac{1}{\pi}\int_{-\infty}^\infty d\omega \frac{\mathcal{P}}{\omega-\omega'}{\rm Re}G^R_{\bf k}(\omega)=-{\rm Im}G^R_{\bf k}(\omega')\,,
\end{equation}
and
\begin{equation}
\frac{1}{\pi}\int_{-\infty}^\infty d\omega \frac{\mathcal{P}}{\omega-\omega'}{\rm Re}\Sigma^R(\omega)=-{\rm Im}\Sigma^R(\omega')\,,
\end{equation}
where the constant term in the real part of the self-energy integrates to zero due to the principal
value integration. Substituting these results into Eq.~(\ref{eq: pop2}), then yields
\begin{eqnarray}
&i&\partial_{t_{ave}}G_{\bf k}^<(t_{ave},t_{rel}=0)= \nonumber\\
&~&\frac{i}{\pi} \int_{-\infty}^\infty d\omega \Biggr \{ {\rm Im}\Sigma^R\left ( \omega\right )G_{\bf k}^<\left (\omega\right )-
{\rm Im}G_{\bf k}^R\left (\omega\right )\Sigma^<\left (\omega\right )\Biggr \}\,,
\end{eqnarray}
because the two terms add together, rather than canceling. But it turns out the integrand exactly vanishes,
which can be seen by examining the Dyson equation for the retarded and the lesser Green's functions in equilibrium:
\begin{equation}
G^R_{\bf k}(\omega)=\frac{1}{\omega+\mu-\epsilon_{\bf k}-\Sigma^R(\omega)}\,,
\end{equation}
and
\begin{equation}
G^<_{\bf k}(\omega)=G^R_{\bf k}(\omega)\Sigma^<(\omega)G^A_{\bf k}(\omega)=|G^R_{\bf k}(\omega)|^2\Sigma^<(\omega)\,.
\end{equation}
So, we find ${\rm Im}G^R_{\bf k}(\omega)={\rm Im}\Sigma^R(\omega)|G^R_{\bf k}(\omega)|^2$,
and hence the integrand vanishes for each $\omega$.

Similar results also hold in the long-time limit if the nonequilibrium Green's functions no longer have average time dependence. This is not so obvious, since the integrals in Eq.~(\ref{eq: pop1}) always integrate over times when the pump is on, and hence they appear to have nontrivial dependence which won't vanish. But in situations where the density of states does not have any singularities, then the Green's functions and self-energies decay exponentially in relative time (for times larger than $2\pi/W$, with $W$ being the bandwidth of the density of states), and so an examination of Eq.~(\ref{eq: pop0}) shows we can limit $\bar t$ such that $t-\bar t < 2\pi/W$, and hence we can restrict the lower limit on the time integral in Eq.~(\ref{eq: pop1}) to
$-2\pi/W$ instead of $-\infty$. Hence, the dependence on the field when $t_{ave}$ becomes large becomes weak. But, for the same reasons why the integral vanished in equilibrium, it will vanish in the long-time case, if the Green's functions take the equilibrium form, but with an average time dependent self-energy (we already know that the retarded Green's function takes this form, but the lesser cannot, or the time derivative of the population will exactly vanish as we showed above).  

This has implications for the exact analysis of the populations. While it may seem like a good approximation to assume the equilibrium form with average time dependent self-energies (and hence an average time-dependent distribution function), such a form cannot hold for the exact solution because the population would not decay in that form, even with the average time dependence of the self-energies and distribution functions. While there are numerous approximations that do this, they are necessarily missing some of the dynamics that occurs for the exact solution to this problem, and it does not appear that this conclusion is well known. In particular, if the approximate result uses the equilibrium form with time-dependent quantities, it cannot describe the exact solution, which brings to question just how accurate can such an approximate solution be?  Put in other words, if the Green's functions and self-energies are given average time dependence, then one needs to start with Eq.~(\ref{eq: pop1}) in order to determine the subsequent dynamics.

Our next step is to extract the long-time behavior of the population dynamics assuming that the average time dependence is weak, so it can be expanded in a Taylor series expansion with respect to average times near $t_{ave}=t$.

\section{Relaxation Dynamics in the Long-Time Limit}

We make an assumption that as we reach the long-time limit, the dependence of quantities on $t_{ave}$ becomes weak, so we can approximate the behavior by the lowest-order terms in a Taylor series about 
$t_{ave}$. In other words, we take quantities like the first term in Eq.~(\ref{eq: pop1}) and expand it in the 
Taylor series
\begin{eqnarray}
&\,&\int_{-\infty}^0 d \bar t \int_{-\infty}^\infty d\omega \int_{-\infty}^\infty d\omega' e^{i(\omega-\omega')\bar t}\nonumber\\
&\,&\Biggr \{  \Sigma^R(t_{ave},\omega)G^<_{\bf k}(t_{ave},\omega')+
\frac{\bar t}{2}\partial_{t_{ave}}\Sigma^R(t_{ave},\omega)G^<_{\bf k}(t_{ave},\omega')\nonumber\\
&\,&+
\Sigma^R(t_{ave},\omega)\frac{\bar t}{2}\partial_{t_{ave}}G^<_{\bf k}(t_{ave},\omega')+\cdots\Biggr \}\,.
\end{eqnarray}
The $\bar t$ terms can be replaced by derivatives with respect to $\omega$ or $\omega'$ of the exponential factor, and then those derivatives can be moved back to the self-energy or Green's function after
integrating by parts. This yields
\begin{eqnarray}
&\,&\int_{-\infty}^0 d \bar t \int_{-\infty}^\infty d\omega \int_{-\infty}^\infty d\omega' e^{i(\omega-\omega')\bar t}\nonumber\\
&\,&\Biggr \{  \Sigma^R(t_{ave},\omega)G^<_{\bf k}(t_{ave},\omega')+
i\partial_{t_{ave}}\partial_\omega\Sigma^R(t_{ave},\omega)G^<_{\bf k}(t_{ave},\omega')\nonumber\\
&\,&-
i\Sigma^R(t_{ave},\omega)\partial_{t_{ave}}\partial_{\omega'}G^<_{\bf k}(t_{ave},\omega')+\cdots\Biggr \}\,.
\end{eqnarray}
Note that in this equation (and subsequent ones), the derivatives act only on the first function to the immediate 
right of the derivative symbol, unless explicitly denoted otherwise by square brackets.
Now, we can integrate over $\bar t$ which produces the same Sokhotskyi-Plemelj-Dirac identity. The delta function piece can next be integrated, as can the principal value integration by employing the appropriate Kramers-Kronig relation. After some significant algebra, one finds the right hand side of the population dynamics becomes
\begin{eqnarray}
&\,&\frac{i}{\pi}\int_{-\infty}^\infty d\omega \Biggr \{ {\rm Im}\Sigma^R(t_{ave},\omega)G^<_{\bf k}(t_{ave},\omega)\nonumber\\
&~&~~~~~~~~~~~~~~~~~~~~~-{\rm Im}G^R_{\bf k}(t_{ave},\omega)\Sigma^<(t_{ave},\omega)\nonumber\\
&\,&~~~~~~~~~~~~+\partial_{t_{ave}}\partial_\omega{\rm Re}\Sigma^R(t_{ave},\omega)G^<_{\bf k}(t_{ave},\omega)\nonumber\\
&\,&~~~~~~~~~~~~-{\rm Re}\Sigma^R(t_{ave},\omega)\partial_{t_{ave}}\partial_\omega G^<_{\bf k}(t_{ave},\omega)\nonumber\\
&\,&~~~~~~~~~~~~-\partial_{t_{ave}}\partial_\omega{\rm Re}G_{\bf k}^R(t_{ave},\omega)\Sigma^<(t_{ave},\omega)\nonumber\\
&\,&~~~~~~~~~~~~+{\rm Re}G_{\bf k}^R(t_{ave},\omega)\partial_{t_{ave}}\partial_\omega \Sigma^<(t_{ave},\omega)\Biggr \}\,.
\label{eq: pop3}
\end{eqnarray}

Eq.~(\ref{eq: gret}) tells us how the retarded Green's function is related to the average time dependent retarded self-energy when the average time dependence is weak. We need a similar result of the lesser Green's function. To work this out, we start with the Dyson equation for the lesser Green's function at $t_{rel}=0$:
\begin{eqnarray}
&~&G_{\bf k}^<(t_{ave},t_{rel}=0)=\nonumber\\
&~&~~~~~~~~~~~~~~~~~\int_{-\infty}^{t_{ave}}d\bar t\int_{-\infty}^{t_{ave}}d\bar t' G^R_{\bf k}\left (\frac{t_{ave}+\bar t}{2},t_{ave}-\bar t\right )\nonumber\\
&~&~~~~~~~~~~~~~~~~~\times\Sigma^<\left ( \frac{\bar t+\bar t'}{2},\bar t -\bar t'\right )G^A_{\bf k}\left (\frac{\bar t'+t_{ave}}{2},\bar t'-t_{ave}\right )\,.
\end{eqnarray}
Since the integrals are dominated by the regions near the upper limits, we let $\bar t,\bar t'\rightarrow \bar t+t_{ave}, \bar t'+t_{ave}$, and then we introduce the Fourier transform with respect to frequency for the relative time dependence in the second argument of each term to yield
\begin{eqnarray}
&~&G_{\bf k}^<(t_{ave},0)=\int_{-\infty}^{0}d\bar t\int_{-\infty}^{0}d\bar t'\frac{1}{8\pi^3}
\int d\omega \int d\omega' \int d\omega''  \nonumber\\
&~&\times e^{i(\omega-\omega')\bar t+(\omega'-\omega'')\bar t'}\Biggr \{ G_{\bf k}^R(t_{ave},\omega)
\Sigma^<(t_{ave},\omega')G^A_{\bf k}(t_{ave},\omega'')\nonumber\\
&~&+\frac{\bar t}{2}\partial_{t_{ave}} G^R_{\bf k}(t_{ave},\omega)\Sigma^<(t_{ave},\omega)G^A_{\bf k}(t_{ave},\omega)
\nonumber\\
&~&
+G^R_{\bf k}(t_{ave},\omega)\frac{\bar t+\bar t'}{2}\partial_{t_{ave}}\Sigma^<(t_{ave},\omega')G^A_{\bf k}(t_{ave},\omega'')\nonumber\\
&~&+G_{\bf k}^R(t_{ave},\omega)\Sigma^<(t_{ave},\omega')\frac{\bar t'}{2}\partial_{t_{ave}}G^A_{\bf k}(t_{ave},\omega'')+\cdots
\Biggr \}\,.
\end{eqnarray}
We only include the lowest-order change with respect to the average time. As done before, the time variables are removed by taking derivatives of the exponentials with respect to the frequencies and then integrating by parts. The integrals over time can then be done, as can the integrals over $\omega$ and $\omega''$ after employing the identity
\begin{equation}
\frac{1}{2\pi}\int d\omega \int_{-\infty}^0 d\bar t e^{i(\omega-\omega')\bar t} F(\omega)=
F(\omega+i0^+)\,.
\end{equation}
Then we find
\begin{eqnarray}
&~&G_{\bf k}^<(t_{ave},0)=\frac{1}{2\pi}
\int d\omega \nonumber\\
&~&\Biggr \{ G_{\bf k}^R(t_{ave},\omega)
\Sigma^<(t_{ave},\omega)G^A_{\bf k}(t_{ave},\omega)\nonumber\\
&~&+\frac{i}{2}\partial_{t_{ave}} \partial_\omega G^R_{\bf k}(t_{ave},\omega)\Sigma^<(t_{ave},\omega)G^A_{\bf k}(t_{ave},\omega)
\nonumber\\
&~&
+\frac{i}{2}\partial_\omega G^R_{\bf k}(t_{ave},\omega)\partial_{t_{ave}}\Sigma^<(t_{ave},\omega)G^A_{\bf k}(t_{ave},\omega)\nonumber\\
&~&-\frac{i}{2}G_{\bf k}^R(t_{ave},\omega)\partial_{t_{ave}}\Sigma^<(t_{ave},\omega)\partial_\omega G^A_{\bf k}(t_{ave},\omega)\nonumber\\
&~&-\frac{i}{2}G_{\bf k}^R(t_{ave},\omega)\Sigma^<(t_{ave},\omega)\partial_{t_{ave}}\partial_\omega G^A_{\bf k}(t_{ave},\omega)+\cdots
\Biggr \}\,.
\end{eqnarray}
The integrand (multiplied by $2\pi$) then gives the Fourier transform of the lesser Green's function. So, we immediately learn that the lesser Green's function (as a function of average time and frequency) behaves like
\begin{eqnarray}
&~&G^<_{\bf k}(t_{ave},\omega)=|G^R_{\bf k}(t_{ave},\omega)|^2\Sigma^<(t_{ave},\omega)\nonumber\\
&~&-{\rm Im}\left [ G_{\bf k}^A(t_{ave},\omega)\partial_{t_{ave}}\partial_\omega G_{\bf k}^R(t_{ave},\omega)\right ] \Sigma^<(t_{ave},\omega)\nonumber\\
&~&-{\rm Im}\left [ G_{\bf k}^A(t_{ave},\omega)\partial_\omega G_{\bf k}^R(t_{ave},\omega)\right ] \partial_{t_{ave}} \Sigma^<(t_{ave},\omega)\,,
\label{eq: glesser}
\end{eqnarray}
when the average time dependence is weak. This is the lesser Green's function analog of Eq.~(\ref{eq: eom_ret}). Unlike in the retarded case, the lesser Green's function does change its 
functional form to first order in the average time dependence.

When we substitute this result into Eq.~(\ref{eq: pop3}), the terms without the derivatives cancel,
as we discussed before, leaving behind the following result for the right hand side of the population dynamics
\begin{eqnarray}
&\,&\frac{i}{\pi}\int_{-\infty}^\infty d\omega \Biggr \{-{\rm Im}[\Sigma^R(t_{ave},\omega)]\nonumber\\
&\,&\times \Biggr ( 
{\rm Im}\left [ G_{\bf k}^A(t_{ave},\omega)\partial_{t_{ave}}\partial_\omega G_{\bf k}^R(t_{ave},\omega)\right ] \Sigma^<(t_{ave},\omega)\nonumber\\
&\,&+{\rm Im}\left [ G_{\bf k}^A(t_{ave},\omega)\partial_\omega G_{\bf k}^R(t_{ave},\omega)\right ] \partial_{t_{ave}} \Sigma^<(t_{ave},\omega)\Biggr )\nonumber\\
&\,&+\partial_{t_{ave}}\partial_\omega{\rm Re}\Sigma^R(t_{ave},\omega)\left [ |G^R_{\bf k}(t_{ave},\omega)|^2\Sigma^<(t_{ave},\omega)\right ]\nonumber\\
&\,&-{\rm Re}\Sigma^R(t_{ave},\omega)\partial_{t_{ave}}\partial_\omega \left [ |G^R_{\bf k}(t_{ave},\omega)|^2\Sigma^<(t_{ave},\omega)\right ]\nonumber\\
&\,&-\partial_{t_{ave}}\partial_\omega{\rm Re}G_{\bf k}^R(t_{ave},\omega)\Sigma^<(t_{ave},\omega)\nonumber\\
&\,&+{\rm Re}G_{\bf k}^R(t_{ave},\omega)\partial_{t_{ave}}\partial_\omega \Sigma^<(t_{ave},\omega)\Biggr \}\,,
\label{eq: pop4}
\end{eqnarray}
to lowest order in the derivatives. It is well known that the retarded Green's function has much weaker average time dependence than the lesser Green's function (we have explicitly verified this in our numerical calculations as well). In particular, after a time on the order of a few $2\pi/W$ past when the field is turned off, the retarded Green's function has essentially reached its long-time limit. We assume we are at times past this limit, so we can neglect all time derivatives with respect to the retarded objects. The remaining derivatives can be evaluated in a straightforward fashion.  After some long algebra, we find
\begin{eqnarray}
&\,&\frac{i}{\pi}\int_{-\infty}^\infty d\omega 
|G_{\bf k}^R(t_{ave},\omega)|^2 \Biggr \{ |G_{\bf k}^R(t_{ave},\omega)|^2\Biggr [ {\rm Im}\Sigma^R(t_{ave},\omega) \nonumber\\
&~&~~~~~~~~~~~~~~~~~~~\times\Big (
\partial_\omega {\rm Im}\Sigma^R(t_{ave},\omega) \big ( \omega+\mu-\epsilon_{\bf k}+{\rm Re}\Sigma^R(t_{ave},\omega)\big )\nonumber\\
&~&~~~~~~~~~~~~~~~~~~~~~~~~~~~+\big ( 1-\partial_\omega {\rm Re}\Sigma^R(t_{ave},\omega)\big ){\rm Im}\Sigma^R(t_{ave},\omega)\Big )\nonumber\\
&~&~~~~~~~~~~~~~~+2{\rm Re}\Sigma^R(t_{ave},\omega)\big ( 1-\partial_\omega\Sigma^R(t_{ave},\omega)\big )\nonumber\\
&~&~~~~~~~~~~~~~~~~\times\big (\omega+\mu-\epsilon_{\bf k}-{\rm Re}\Sigma^R(t_{ave},\omega)\big ) \Biggr ] \partial_{t_{ave}}\Sigma^<(t_{ave},\omega)\nonumber\\
&~&~~~+\big (\omega+\mu-\epsilon_{\bf k}-{\rm Re}\Sigma^R(t_{ave},\omega)\big )
\partial_\omega \partial_{t_{ave}} \Sigma^<(t_{ave},\omega)\Biggr \}\,.
\label{eq: pop_final}
\end{eqnarray}
This looks like a rather formidable result, but it actually isn't so bad. If the lesser self-energy relaxes 
exponentially (which is not obvious from these results, but often occurs numerically), then we can approximate $\partial_{t_{ave}} \Sigma^<(t_{ave},\omega)\approx -\Gamma \delta \Sigma^<(t_{ave},\omega)$. Here $\delta \Sigma^<(t_{ave},\omega) = \Sigma^<(t_{ave},\omega)-\Sigma^<(\infty,\omega)$. Then, because of $|G^R_{\bf k}|^2$ is sharply peaked around $\omega\approx\epsilon({\bf k})$ and the fact that the remaining factors are of order 1, we find the population will also relax exponentially. Now this is not a proof that the system will always
relax exponentially, but it does show that it is consistent with such behavior.

\begin{figure}
  \includegraphics[width=\columnwidth]{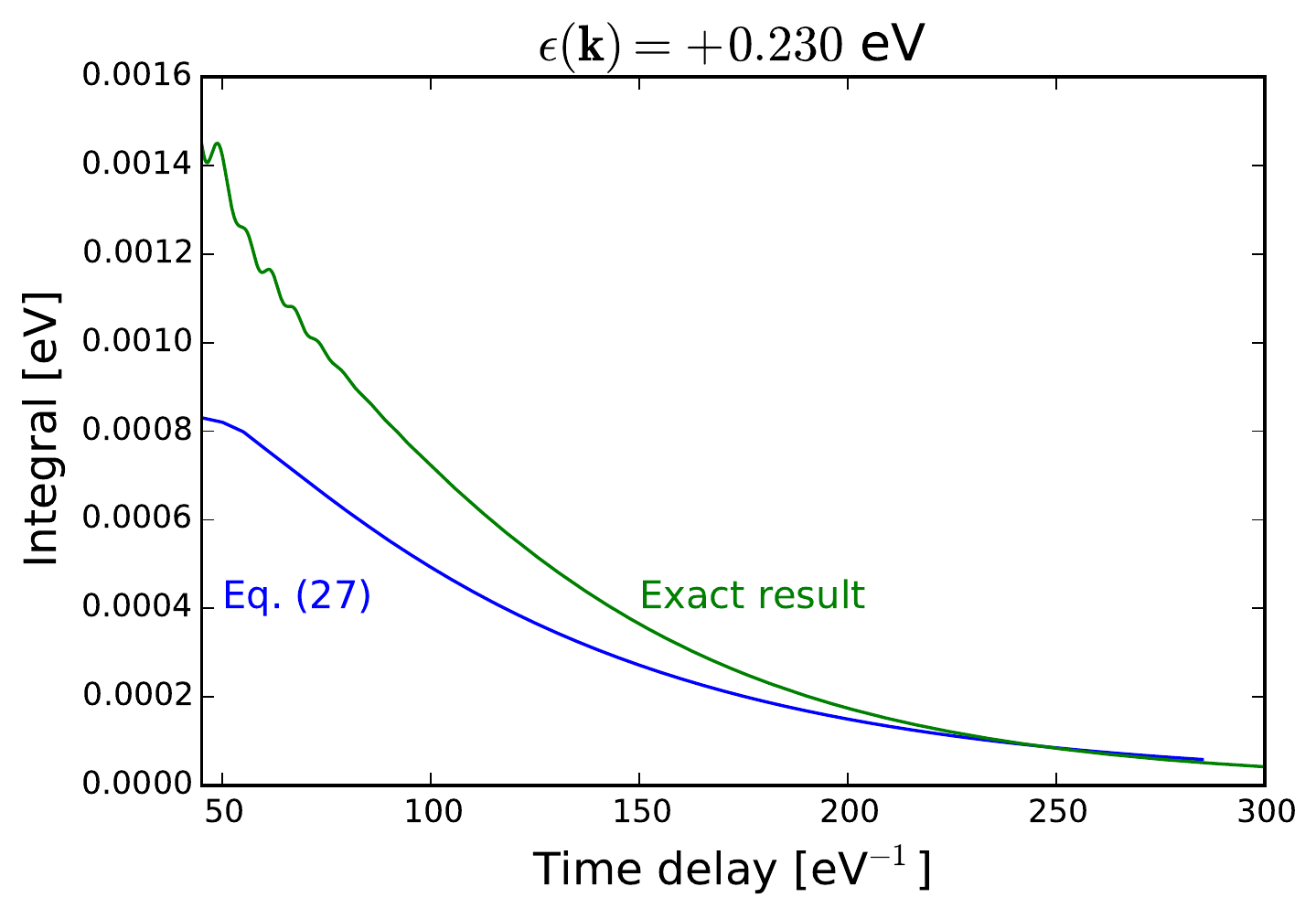}
  \caption{\label{fig: total_above}\col 
    Exact total scattering integral compared to the approximate result in Eq.~(\ref{eq: pop_final}), for $\Omega=0.2t_{hop}$ and
two-d noninteracting bandstructure $\epsilon({\bf k})=0.23$~eV.  Note how the two curves agree as $t\rightarrow\infty$.}
\end{figure}

\begin{figure}
  \includegraphics[width=\columnwidth]{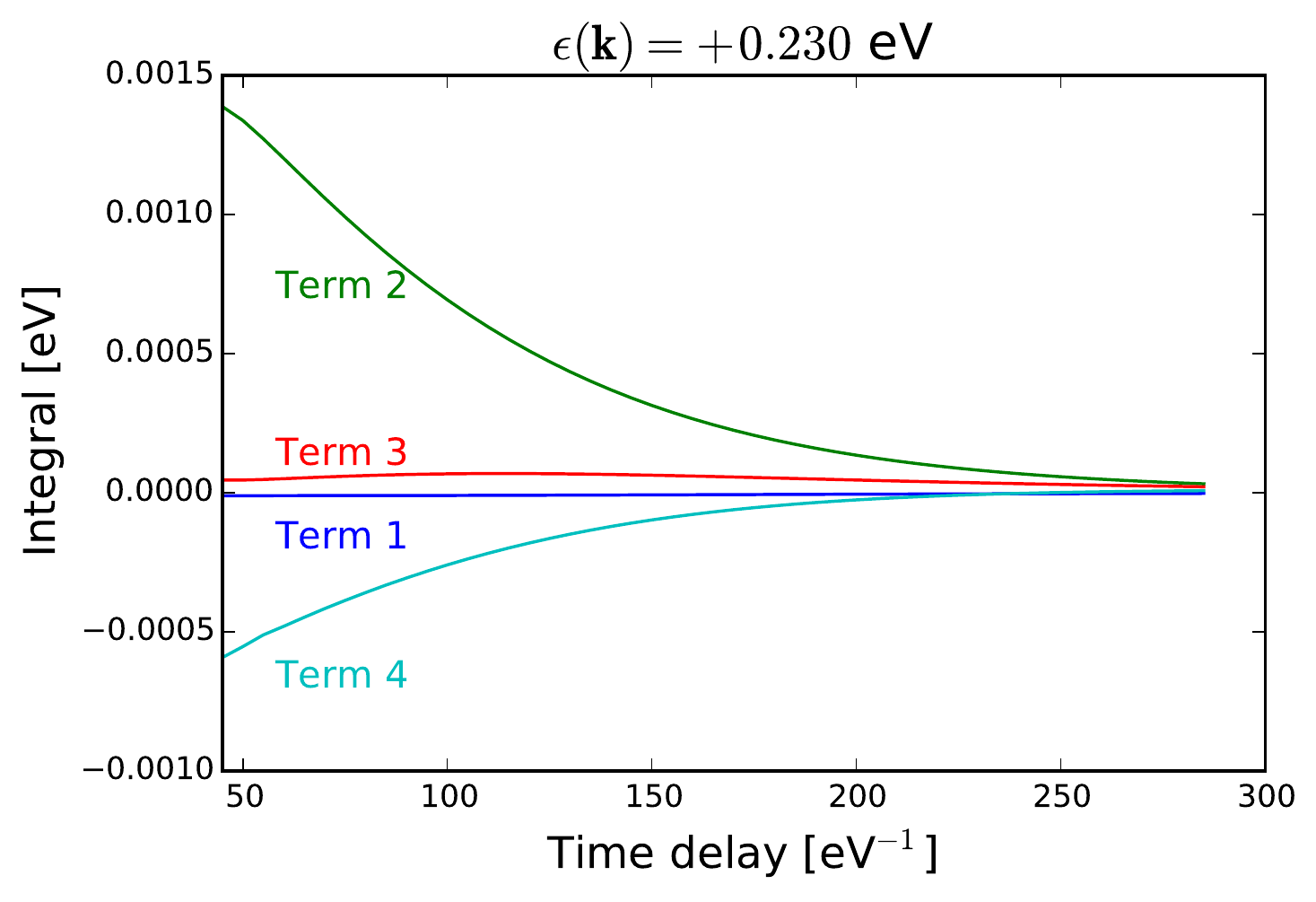}
  \caption{\label{fig: terms_above}\col 
    Plot of each of the four terms in Eq.~(\ref{eq: pop_final}), for $\Omega=0.2t_{hop}$ and
two-d noninteracting bandstructure $\epsilon({\bf k})=0.23$~eV.  }
\end{figure}

\begin{figure}
  \includegraphics[width=\columnwidth]{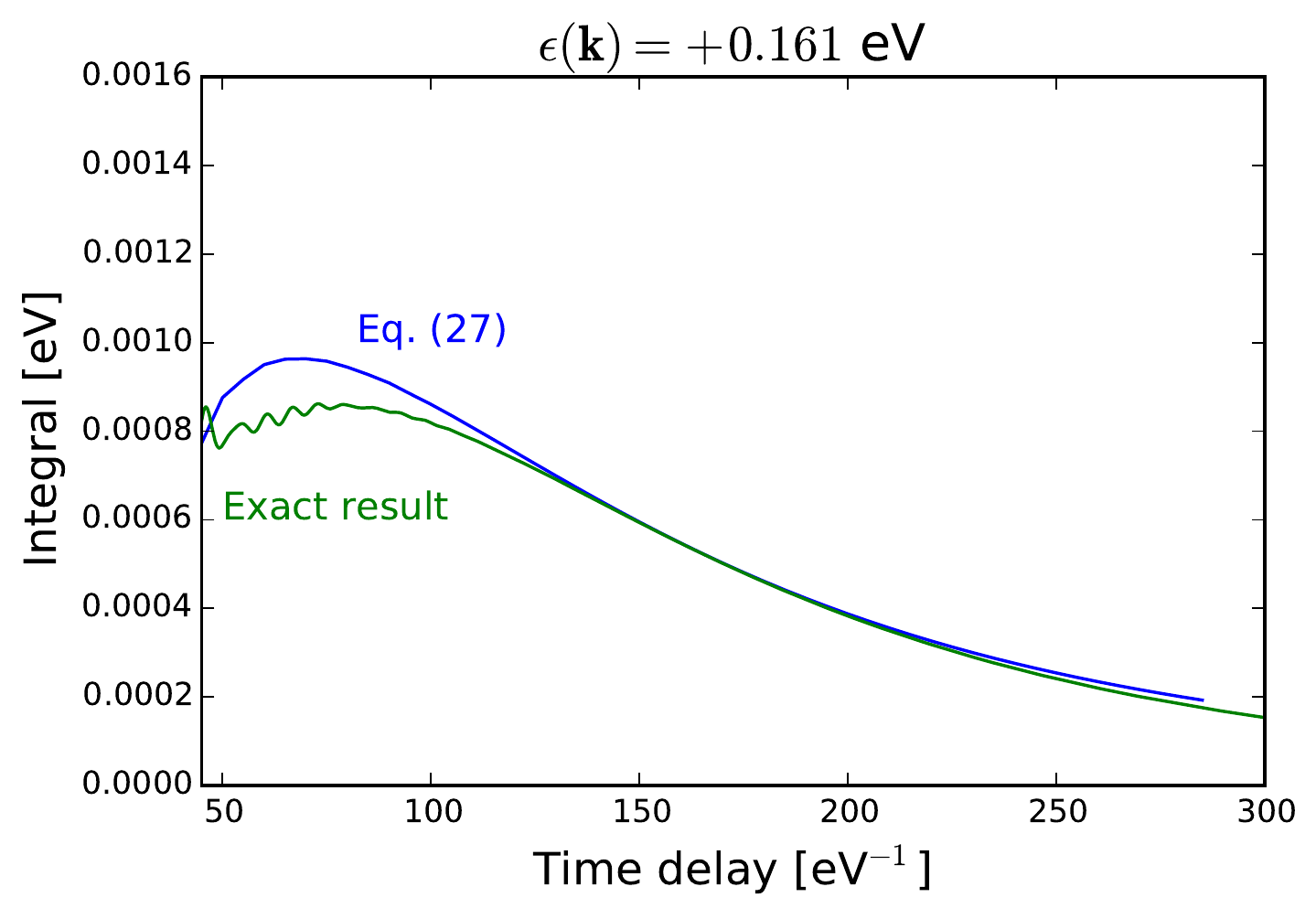}
  \caption{\label{fig: total_below}\col 
   Exact total scattering integral compared to the approximate result  in Eq.~(\ref{eq: pop_final}), for $\Omega=0.2t_{hop}$ and
two-d noninteracting bandstructure $\epsilon({\bf k})=0.156$~eV. Note how the two curves agree at long times.}
\end{figure}

\begin{figure}
  \includegraphics[width=\columnwidth]{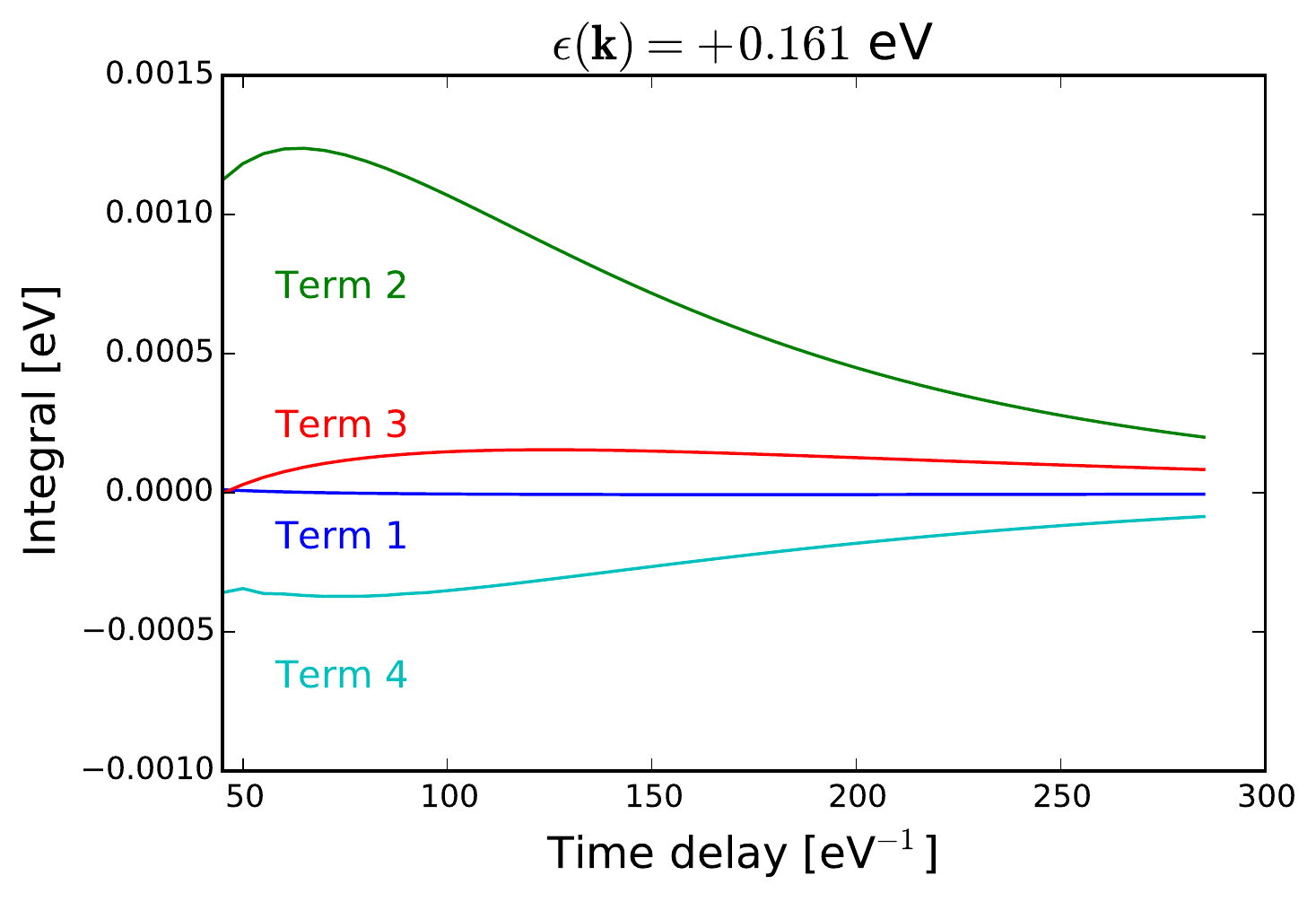}
  \caption{\label{fig: terms_below}\col 
    Plot of each of the four terms in Eq.~(\ref{eq: pop_final}), for $\Omega=0.2t_{hop}$ and
two-d noninteracting bandstructure $\epsilon({\bf k})=0.156$~eV. }
\end{figure}

In order to gain some more insight into Eq.~(\ref{eq: pop_final}), we have run simulations on the Holstein model on a two-dimensional lattice, incorporating Migdal -Eliashberg theory for the electron self-energy, but not self-consistently updating the phonons~\cite{lex_holstein_noneq}; this can be viewed as an
application of nonequilibrium dynamical mean-field theory~\cite{freericks_nedmft,eckstein_review}. The phonon energy is chosen to be $0.2t_{hop}$, with $t_{hop}$ the hopping integral (not to be confused with the time), which we set equal to 1~eV; we also add a small-weight scatterer with a frequency around 0.01~eV to ensure relaxation for low-energy electrons~\cite{rameau}. We first plot the exact total scattering integral versus the approximate result in Eq.~(\ref{eq: pop_final}) in Fig.~\ref{fig: total_above}. Note how the two results agree at long times (the small differences are most likely due to numerical issues associated with taking numerical derivatives). We plot the results for the four different terms [separated by the plus signs in the large curly brackets in Eq.~(\ref{eq: pop_final})] and denoted Term 1 to 4. Fig.~\ref{fig: total_below} plots the comparison of the exact and approximate scattering integrals for the case where the electron's
excitation energy lies above the Fermi surface but is less than the phonon energy. Note again how they match at long times.  Fig.~\ref{fig: terms_above} plots the different contributions to the sum for an electron at a momentum {\bf k}, which lies more than the phonon energy above the chemical potential, while Fig.~\ref{fig: terms_below} plots these results for an electron at a momentum {\bf k} which has an energy above the Fermi energy, but within the phonon window. What we see is that there isn't one term which is always dominant and the terms can enter with opposite signs. After the cancellations, the majority of the signal comes from Term 2 alone. Nevertheless, this makes it more difficult to obtain analytic estimates of the relaxation.

We also investigate the distribution functions of the system. Loosely speaking, one can define the 
distribution function via the ratio of the imaginary part of a lesser object divided by the imaginary part of the corresponding retarded object. For example, we can define $f_G(t_{ave},\omega)=-{\rm Im}G^<(t_{ave},\omega)/[2{\rm Im}G^R(t_{ave},\omega)]$, which would be the Fermi-Dirac distribution
in equilibrium due to the fluctuation-dissipation theorem. Similarly, we can define a self-energy distribution function via $f_{\Sigma}(t_{ave},(\omega)=-{\rm Im}\Sigma^<(t_{ave},\omega)/[2{\rm Im}\Sigma^R(t_{ave},\omega)]$. As we have determined above, one must have $f_G\ne f_{\Sigma}$ during the relaxation process, because once they become equal, regardless of the specific functional form of $f_G(t_{ave},\omega)$, the time derivative of the population vanishes.

\begin{figure*}
  \includegraphics[width=6.8in]{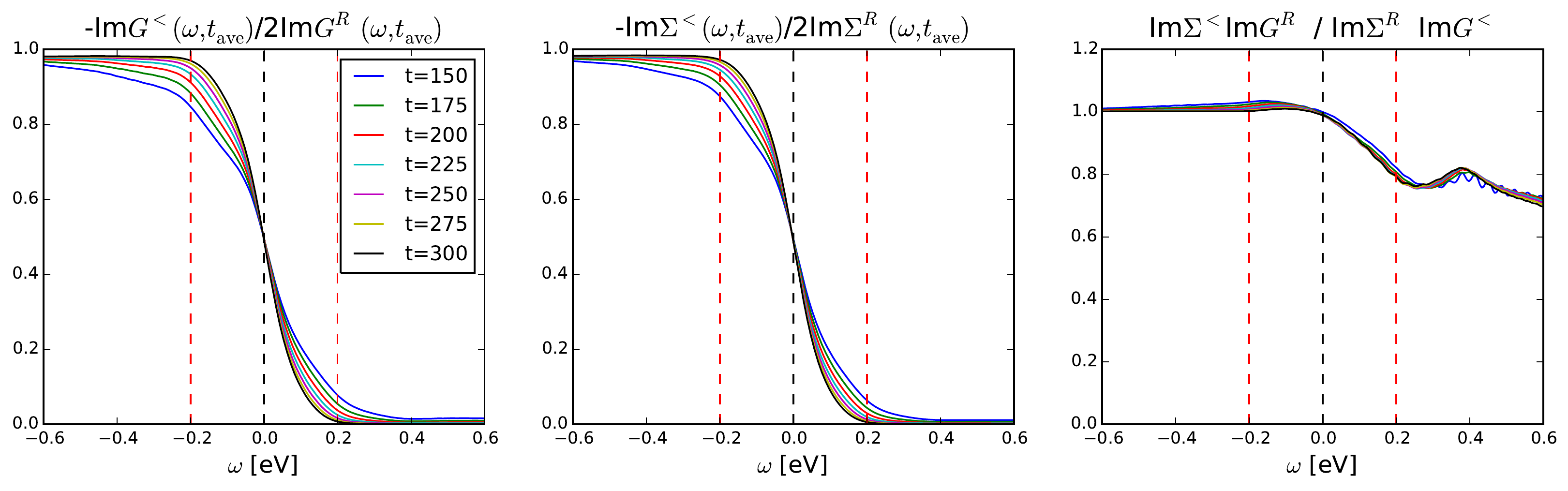}
  \caption{\label{fig: distribution}\col 
    Plot of the effective distribution functions $f_G$ (left) and $f_{\Sigma}$ (center) and their ratio (right) for the local Green's function and self-energy. }
\end{figure*}

These two distribution functions and their ratio are plotted in Fig.~\ref{fig: distribution}, where we see that both the Green's function
distribution and the self-energy distribution are approaching Fermi-Dirac distributions, but they do so at
different rates, so their ratio is not precisely one (with the largest deviations being at high energy, where the smallness of the distribution function enhances the ratio, and where we expect the deviations to be large until those high-energy electrons fully relax). 

What relevance does this have for approximate models and phenomenology? If we define the distribution functions as we have, with time-dependent functions of frequency multiplying the spectral functions, then we find that one can never assume that they precisely take the quasi-equilibrium form of a high-temperature Fermi-Dirac distribution, because to do so would imply that there is no further relaxation. This is a general statement that always holds, but particularly so for systems that one might want to describe via a hot-electron model, where the electrons have a quasi-equilibrium distribution due to fast equilibration via electron-electron scattering and then a subsequent longer-time cooling via interaction with the phonons.
In fact, many-body physics generically forbids this from occurring---if we use a Fermi-Dirac
distribution for the electronic Green's function, then calculating the self-energy will not yield the same Fermi-Dirac distribution for the lesser component if the phonons are at a different temperature (within Migdal-Eliashberg theory). While the general phenomenology of this behavior is expected for many systems, the analysis given here shows that the distributions can never be fully quasi-equilibrium, and they must evolve as functions of time. In particular, one need not even have the same distribution function for the Green's function and for the self-energy!

One of the key unsolved questions that remains is how does the system determine the way in which it relaxes (exponentially or some other way) and what governs the exponential decay rate (if it is exponential decay)? Some progress on this question has been made recently. In the case of weak-coupling, where the self-energy is not updated self-consistently, one can show it is the imaginary part of the self-energy that governs this relaxation~\cite{prx}, while in self-consistent theories, the relaxation remains quite exponential in its functional form, and the decay rate is given semiquantitatively by the imaginary part of the self-energy, there is no explicit derivation of how to find this dependence~\cite{entropy}. In particular, the separation of the system into the terms given by the lesser Green's function and the lesser self-energy in the scattering integrals show that the true decay rate is more than an order of magnitude smaller than either of the scattering integrals due to their near cancellation. In this work, we showed how one can extract an explicit formula for the decay rate of the system, but it cannot be evaluated, or even approximated analytically, because we cannot easily estimate derivatives with respect to time. Hopefully further work in this area will shed light onto this interesting and important question.

\section{Conclusions}

We have analyzed the scattering integrals for the nonequilibrium many-body problem. We show how
the total vanishes when there is no average time dependence and we show what the leading-order
behavior is for the long-time limit where the scattering rate has little time dependence, and the time-dependence of most quantities is weak. These results are all exact, and independent of any Hamiltonian
or scattering mechanism. Unfortunately, they do not show that the relaxation is generically exponential,
nor do they show that the relaxation rate is given by any simple result related to the retarded self-energy.
The behavior is more complex, and cannot be explained in such simple terms. Nevertheless, we do
have the appropriate limiting forms for the scattering integrals which provide the exact relaxation
in the long-time limit. Since these are exact results, they are a better starting point to consider than
other approximate approaches, which are likely to violate exact relations of the full solution to the
problem. Working out a systematic approximation scheme based on this approach, or finding an
appropriate analysis that allows one to make analytic predictions of the relaxation behavior, are
both topics for future study of this problem.

\begin{acknowledgements}
This work was supported at Georgetown by the Department of Energy, Basic Energy Sciences, under grant number DE-FG02-08ER46542. H.R.K. was supported by the Department of Science and Technology in India. J.K.F. was also supported by the McDevitt bequest at Georgetown University. 
\end{acknowledgements}

\end{document}